\documentclass{aa}
\usepackage{astron,graphics}

\renewcommand\d{\mathrm{d}}

\begin{document}

\thesaurus{12(12.03.4;  
              12.12.1)} 
    
\title{A New Crystallographic Method for Detecting Space Topology}

\author{J.-Ph. Uzan\inst{1,2} \and R. Lehoucq\inst{3} \and J.-P. Luminet\inst{2}}

\offprints{J.-Ph. Uzan}

\institute{
D\'epartement de Physique Th\'eorique, Universit\'e de Gen\`eve,
24 quai E. Ansermet, CH-1211 Geneva, Switzerland\\
email: uzan@amorgos.unige.ch
\and
D\'epartement d'Astrophysique Relativiste et de Cosmologie,
Observatoire de Paris, CNRS--UPR 176, F-92195 Meudon, France\\
email: Jean-Pierre.Luminet@obspm.fr
\and
CE-Saclay, DSM/DAPNIA/Service d'Astrophysique, 
F-91191 Gif sur Yvette cedex, France\\
email: roller@discovery.saclay.cea.fr
}

\titlerunning{A new method for detecting space topology}


\maketitle

\begin{abstract}
Multi--connected universe models with space identification scales
smaller than the size of the observable universe produce topological
images of cosmic sources.  We generalise to locally hyperbolic spaces
the crystallographic method, aimed to detect the topology from
three--dimensional catalogs of cosmological objects. Our new \- method
is based on the construction of a {\it collecting--cor\-related--pair}
technique which enhances the topological signature and can make it
detectable. The main idea is that in multi--connected universes, equal
distances occur more often than by chance. We present an idealised
version of this method as well as numerical simulations, we discuss
the statistical relevance of the expected signature and we show how
the extraction of a topological signal may also lead to a precise
determination of the cosmological parameters.  A more realistic
version of the method which takes account of uncertainties in the
position and velocity data is then discussed.  We apply our technique
to presently available data, namely a quasar catalog, and discuss the
significance of the result.  We show how the improved crystallographic
method both competes with the two--dimensional methods based on cosmic
microwave background analyses, and suffers from the same drawback: the
shortage of present observational data.

\vskip0.2cm
\noindent{\bf Preprint numbers:} UGVA-DPT 1999/03--1028. DARC 99-03
\end{abstract}


\section{Introduction}
\label{I}

Recently there have been many advances in the development of methods
to detect or constrain the topology of the universe (i.e. of its
spatial sections).  Since the revival of cosmic topology (see
Lachi\`eze-Rey and Luminet \cite*{lachieze95} and the proceedings of
the workshop {\it Cosmology and topology} \cite*{cqg98} for the latest
developments) many methods using either the cosmic microwave
background (see e.g. Stevens et al.  \cite*{stevens93}, Cornish et al.
\cite*{cornish98:a,cornish98:b,cornish98:c}, Levin et al.
\cite*{levin97,levin98} and Uzan et al.  \cite*{uzan98:a,uzan98:b}),
or discrete sources such as clusters, quasars, etc. (see e.g. Lehoucq
et al.  \cite*{lehoucq96} and Roukema and Edge \cite*{roukema97}) were
investigated.  The latter class of methods is based on the fact that
an observer in a multi-connected universe is looking at a finite part
of its universal covering space \cite{ellis71}.  We thus expect to see
multiple images of any object \cite{lachieze95}.  This property was
first applied to specific objects to provide bounds on the size of the
universe using our Galaxy \cite{sokolov71,fagundes87}, the Coma
cluster \cite{gott80,roukema96} and quasars \cite{roukema97}.  It was
then applied statistically through the so--called crystallographic
method \cite{lehoucq96}, the applicability of which was discussed by
Fagundes and Gausmann \cite*{fagundes97} in Euclidean manifolds and by
Lehoucq et al.  \cite*{lehoucq99} in hyperbolic and elliptic
manifolds, see also Gomero et al.  \cite*{gomero98}, Fagundes and
Gausmann \cite*{fagundes98}. The crystallographic method is a
statistical one which was framed as to extract characteristic
lengths associated with the topology of the universe, and thus which
gets around the problem of recognizing objects at different ages and
orientations \cite{lehoucq96}.

Today, multi-connected universes with locally Euclidean spatial 
sections are well constrained both by the absence of a cut-off in the 
angular power spectrum of the cosmic microwave background temperature 
anisotropies \cite{stevens93}, and by the crystallographic method 
\cite{lehoucq96,texas98} which puts a lower bound on the 
characteristic size, $L$ say, of Euclidean space models to $L\geq 
3000\,h^{-1}\hbox{Mpc}$ (with $h=H_0/100\, 
\hbox{km\,s}^{-1}\hbox{Mpc}^{-1}$, $H_0$ being the Hubble constant).  
But there is still no reliable constraint on a universe with locally 
hyperbolic spatial sections either from the cosmic microwave 
background methods \cite{inoue98,bond98} or from the clusters and 
quasars catalogs \cite{lehoucq99}.  Thus, the quest for cosmic 
topology will probably be longer than expected.

The goal of this article is to improve the crystallographic method so
as to make it suitable for the detection of topology in locally
hyperbolic spaces.  For that purpose, we first review the
crystallographic method in Sect.~\ref{II}, comment on its
applicability and also on recent attempts at its improvement and
generalisation.  This leads us to propose a new method to
detect the topology using a catalog of discrete sources
(Sect.~\ref{III}), aimed to enhance the topological signal by
collecting all correlated pairs of images.  We first apply the
technique to simulated catalogs (Sect.~\ref{IV}) in order to study its
statistical relevance, then to a real catalog of more than $11,000$
quasars (Sect.~\ref{V}).  We discuss this result after having studied
all the uncertainties in the implementation of the method and in
real data.

\section{Ups and downs of the crystallographic method}
\label{II}

Assuming that the four--dimensional manifold ${\cal M}$ describing the 
universe is globally hyperbolic \cite{hawking73}, which is the case 
since we reduce our study to universes which can locally be described 
by a Friedmann-Lema\^{\i}tre spacetime, it can be split as ${\cal 
M}=R\times\Sigma$ where the spatial sections $\Sigma$ are locally 
homogeneous and isotropic.

Now, if the universe is multi-connected, $\Sigma$ can be written as
\begin{equation}
\Sigma\equiv  X/\Gamma,
\end{equation}
where $X$ is either the Euclidean space ($E^3$), the spherical space 
($S^3$) or the hyperbolic space ($H^3$), and $\Gamma$ is the holonomy 
group, namely a finite discrete subgroup without fixed point of the 
full isometry group of $X$ \cite{lachieze95}.  The determination of 
the universal covering $X$ will be given by the knowledge of the 
cosmological parameters such as the density parameter $\Omega_0$ and 
the cosmological constant $\Lambda$, which will help us to determine 
the local geometry (since Einstein's equations are partial 
differential equations, there is locally no difference between a 
multi-- and a simply--connected universe).

The three--dimensional manifold $\Sigma$ can be conveniently described 
by its fundamental domain, which is a polyhedron with $2K$ faces 
identified by pairs through the elements $g$ of the holonomy group 
$\Gamma$ \cite{lachieze95}.  A classification of these manifolds can 
be found in Lachi\`eze-Rey and Luminet \cite*{lachieze95} for Euclidean 
and elliptic manifolds, and by using the software {\it SnapPea} 
({\tt http://www.northnet.org/weeks}) for hyperbolic manifolds, see 
also Thurston \cite*{thurston79} and Wolf \cite*{wolf84}.

The crystallographic method \cite{lehoucq96} relies on a property of
multi--connected universes according to which each topological image
of a given object is linked to the other one by the holonomies of
space which are indeed unknown as far as the topology is not
determined.  The only thing we know is that these holonomies are
isometries.  For instance in a 3--torus universe, to each
holonomy $g$ is associated a distance $\lambda_{g}$, equal to the
length of the translation by which the fundamental domain is moved to
produce the tessellation in the covering space.  Assuming the
fundamental domain contains $N$ objects (e.g. galaxy clusters), if we
calculate the mutual 3D--sepa\-rations between every pair of topological
images (inside the particle horizon), the distances $\lambda_{g}$ will
occur at least $N$ times for each copy of the fundamental domain, and
all other distances will be spread in a smooth way between zero and
two times the horizon distance.  In a pair separation histogram which
plots the number of pairs versus their 3D--separations, the repetition
scales $\lambda_{g}$ should thus produce spikes.  Simulations in
locally Euclidean manifolds showed that the pairs between two
topological images of the same object drastically emerge from ordinary
pairs \cite{lehoucq96} in the histogram.

As it was recently explained \cite{lehoucq99}, two kinds of pairs can
create a spike, namely (we keep the notation and vocabulary introduced
in Lehoucq et al. \cite*{lehoucq99})
\begin{enumerate}
\item {\it Type I pairs} of the form $\lbrace g(x),g(y)\rbrace$, since

$\hbox{dist}[g(x),g(y)]=\hbox{dist}[x,y]$ for all points and all 
elements $g$ of $\Gamma$.

\item {\it Type II pairs} of the form $\lbrace x,g(x)\rbrace$ if 
$\hbox{dist}[x,g(x)]=\hbox{dist}[y,g(y)]$ for at least some points and 
elements $g$ of $\Gamma$.
\end{enumerate}
It has been shown
\cite{lehoucq99,gomero98} that in hyperbolic manifolds, type
II pairs will
not exist since the holomies are not Clifford translations
(i.e. they are
such that $\hbox{dist}[x,g(x)]$ depends on $x$).

Indeed, type I pairs are always present whatever the topology but 
their number is roughly equal to the number of copies of the 
fundamental polyhedron within the catalog's limit.  This number is too 
small to create a spike in the pair separation histogram 
\cite{lehoucq99} and depends on the cosmological parameters since a 
catalog of redshift depth $z$ will contain all the objects located 
within the 2--sphere centered on the observer and of radius $\chi(z)$, 
the radial distance obtained by integration of the null geodesic 
equation up to a redshift $z$.

Two possible solutions were proposed to improve the crystallographic 
method in order to detect the topology and thus to extract the 
signature of type I pairs in catalogs.
\begin{enumerate}
\item Fagundes and Gaussmann \cite*{fagundes98} developed a variant of 
the crystallographic method where they subtract the pair separation 
histogram for a simulated catalog in a simply-connected universe from 
the observed pair separation histogram (with the same number of 
objects and the same cosmological parameters).  The result is ``a plot 
with much oscillation on small scale, modulated by a long wavelength 
quasi-sinusoidal pattern''.  The statistical relevance of this 
signature still has to be investigated.
\item Gomero et al. \cite*{gomero98} proposed splitting the catalog
into ``smaller'' catalogs and averaging the pair separation histograms
built from each sub--catalog to reduce the statistical noise and
extract a topological signal from what they call ``the
non-translational isometries''.  The feasability of this method has
not yet been demonstrated.
\end{enumerate}

We now develop a new method based on the property that whatever the
topology, type I pairs will always exist.  Indeed, as we have shown in
Lehoucq et al. \cite{lehoucq99}, this does not lead to any observable
spike in the pair separation histogram.  We thus have to improve the
crystallographic method to enhance the topological signature by
``collecting'' all these correlated pairs together.  This can be
achieved by defining what we call a {\it collecting correlated pair}
method (hereafter CCP--method).  Such an approach (described in
detail in the next section) will not be able to determine the exact
topology but will give a signature of the existence of at least one
compact spatial dimension on sub--horizon scale, which is indeed a
first step toward the determination of the topological scale of the
universe.  Once such a signature is obtained, many routes are possible
to determine the exact topology.  We shall briefly refer to them in
the final discussion.

\section{How can we use a 3D-catalog to detect the topology of the universe~?}
\label{III}

\subsection{Basic idea}

As stressed in the former paragraph, type I pairs will always exist in
multi--connected spaces as soon as one of the characteristic lengths
of the fundamental domain, such as the injectivity radius $r_{\rm
inj}$ (see, e.g., fig.  10 in Luminet and Roukema \cite*{cargese98}),
is smaller than the Hubble scale.  Defining $\left.g_i\right|_{1\leq
i\leq 2K}$ as the $2K$ generators of $\Gamma$ and referring to $x$ as
the position of the image in the universal covering space $X$, we
have:
\begin{enumerate}
\item $\forall\, x,y\in X$, $\forall g\in\Gamma$,
\begin{eqnarray}
\hbox{dist}[g(x),g(y)]&=&\hbox{dist}[x,y].\label{zero1}
\end{eqnarray}
We will refer to these pairs as $xy$--pairs.

\item $\forall\, x\in X$, $\forall g_1,g_2\in\Gamma$,
\begin{eqnarray}
\hbox{dist}[g_1(x),g_1\circ 
g_2(x)]&=&\hbox{dist}[x,g_2(x)].\label{zero2}
\end{eqnarray}
We will refer to these pairs as $xg(x)$-pairs.
\end{enumerate}
Both the $xy$--pairs and the $xg(x)$-pairs are type I pairs.

To collect all these pairs and enhance the topological signal, we 
define the {\it CCP--index} of a catalog containing $N$ objects as 
follows.
\begin{enumerate}
\item We compute all the 3D--distances $\hbox{dist}[x,y]$ for all 
pairs of points within the catalog's limit.

\item We order all these distances in a list $\left.d_i\right|_{1\leq 
i\leq P}$, where $P\equiv N(N-1)/2$ is the number of pairs, such that 
$d_{i+1}\geq d_i$.

\item We create a new list of increments defined by
\begin{equation}
\forall i\in [1...P-1],\quad \Delta_i\equiv d_{i+1}-d_i \label{incr}
\end{equation}
(keeping all the equal distances, if any, in the list).

\item We then define the CCP--index ${\cal R}$ as
\begin{equation}
{\cal R} \equiv \frac{{\cal N}}{P-1} \label{CCP},
\end{equation}
where ${\cal N} \equiv \hbox{card}(\lbrace i, \Delta_i=0 \rbrace)$, so 
that $0 \leq {\cal R} \leq 1$.
\end{enumerate}

With such a procedure, all type I pairs will contribute to ${\cal N}$.  
For instance, if a given distance appears 4 times in the list 
$\left.d_i\right|_{1\leq i\leq P}$, it will contribute to 3 counts in 
${\cal N}$.  Compared to the former crystallographic method, all the 
correlated pairs are gathered into a single spike, instead of being 
diluted into the noise of the histogram pair separation.

Indeed, in a more realistic situation, one has to take into account 
bins of finite width $\epsilon$ and replace ${\cal N}$ by
\begin{eqnarray}
{\cal N}_\epsilon&\equiv& \hbox{card}(\lbrace i, 
\Delta_i\in[0,\epsilon[ \rbrace)\label{bin}
\end{eqnarray}
in the computation of ${\cal R}$.  The effect of such a binning will 
be discussed in Sect.~\ref{V}.  We now focus on the ``idealised'' 
version of the procedure and study the magnitude of the CCP--index in 
multi-connected models.

For that purpose, let us assume that the catalog is obtained from an 
initial set of $A$ objects lying in the fundamental domain and that 
there are $B$ copies of this domain within the catalog's limits 
($B=0$ if the whole observable universe up to the catalog's limit is 
included inside the fundamental domain).  The total number of images is 
$N=A(B+1)$.  Indeed $B$ is not usually an integer but we assume it is, 
in order to {\it estimate} the amplitude of ${\cal R}$ and compare it 
with the result in a simply-connected model.

We now need to count the number of copies of a given pair.
\begin{enumerate}
\item Each $xy$--pair will be represented $B+1$ times, so that the 
$A(A-1)/2$ pairs contribute to $BA(A-1)/2$ counts in ${\cal N}$.  

\item For $xg(x)$-pairs, let $\nu_1(\Sigma,B)$ be the contribution to 
${\cal N}$ when $A = 1$.  For all $x\in\Sigma$, the total contribution 
of the $xg(x)$-pairs is then $A^2 \nu_1(\Sigma,B)$.
$\nu_1(\Sigma,B)$ can be computed by considering a catalog generated
from an initial set of $A=1$ object and has the property that whatever
the manifold $\Sigma$, $\nu_1(\Sigma,0)=0$ and $\nu_1(\Sigma,B)$ is an
increasing function of $B$.

\item Summing these two type I pair contributions we obtain:
\begin{equation}
{\cal N}_{min}=A\left[(A-1)\frac{B}{2}+A \nu_1(\Sigma,B)\right].
\label{amplitude}
\end{equation}
\end{enumerate}

Indeed, if the holonomy group $\Gamma$ contains Clifford translations 
allowing for type II pairs, or if there are ``fake'' pairs (i.e. such 
that $\hbox{dist}[x,y]=\hbox{dist}[u,v]$), ${\cal N}_{min}$ computed 
from (\ref{amplitude}) will give a lower bound for the true ${\cal 
N}$.

The normalised CCP--index (\ref{CCP}) follows straightforwardly.
${\cal R}$ is a good index for extracting the topological signal since
\begin{enumerate}
\item when $B = 0$ (i.e. when the fundamental
domain is greater than the catalog's spatial scale), ${\cal R} = 0$,

\item when the number of sources in the fundamental domain becomes 
important, it behaves as
\begin{equation}
{\cal R} \rightarrow\frac{B+2\nu_1(\Sigma,B)}{(B+1)^2}\quad 
\hbox{as}\quad A\rightarrow\infty.
\end{equation}
\end{enumerate}

\subsection{Application to a 2--dimensional flat manifold}

As an illustrating example, we perform the computation in 
2--dimensional flat manifolds, and more particularly in the case of 
the 2--torus $T^2$.  Such a space contains Clifford translations, 
however we count only the contribution of type I pairs to ${\cal R}$.  
In that case, the only quantity which needs to be computed is 
$\nu_1(T^2,B)$.

\begin{figure}
\resizebox{\hsize}{!}{\includegraphics{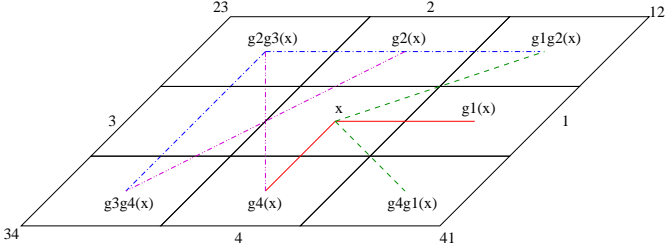}}
\caption{Example of the 2--torus with $B=8$. We have
represented the different type I pairs that contribute to
$\nu_1(T^2,8)$.}
\label{fig2}
\end{figure}

Consider the 2--torus $T^2$ with its eight nearest neighbours ($B=8$)
[see Fig.~\ref{fig2} for the notation].  The pairs
\begin{itemize}
\item $(x,g_1(x))$ and $(x,g_4(x))$ appear 6
times,
\item $(x,g_1\circ g_2(x))$ and $(x,g_1\circ g_4(x))$ appear 4
times,
\item $(g_2\circ g_3(x),g_1\circ g_2(x))$ and
$(g_2\circ
g_3(x),g_3\circ g_4(x))$ appear 3 times,
\item $(g_4(x),g_2\circ g_3(x))$,
$(g_4(x),g_1\circ g_2(x))$,
$(g_4(x),g_2\circ g_3(x))$, $(g_3(x),g_1\circ
g_2(x))$
and $(g_3(x),g_1\circ g_4(x))$
appear 2 times,
\end{itemize}
so that $\nu_1(T^2,8)=2\times5+2\times3+2\times2+4\times1=24$.  This 
leads to ${\cal N}=4A(7A-1)$ whereas $P=9A(9A-1)/2$ so that
\begin{equation}
\frac{24}{35}\leq {\cal R} \leq\frac{56}{81},\quad\forall\, A\geq1.
\end{equation}

Using the same technique one could compute the same quantities 
for various 2--dimensional topologies such as hyperbolic $n$--holes 
tori.

\subsection{Application to 3--dimensional manifolds}

\begin{figure}
\resizebox{\hsize}{!}{\includegraphics{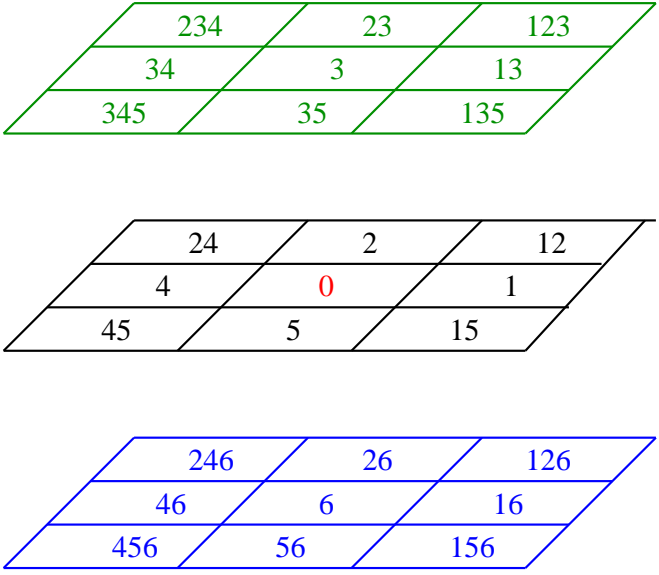}}
\caption{Example of the 3--torus with $B=26$, convention for labelling 
the copies of the fundamental domain.  The image of a point in the 
fundamental domain (at the center) is mapped into the cell e.g. $123$ 
by $g_1\circ g_2\circ g_3$.}
\label{fig3}
\end{figure}

\begin{figure}
\resizebox{\hsize}{!}{\includegraphics{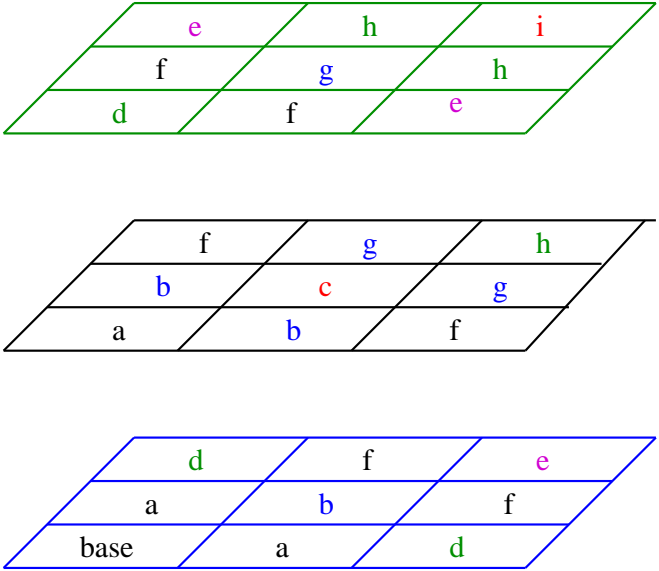}} 
\caption{Example of the 3--torus. The different families of pairs that 
contribute to $\nu_1(T^3,26)$ as labelled in the text starting from the 
base point in $456$.}
\label{fig3b}
\end{figure}

In the case of 3--dimensional manifolds, the
computation of
$\nu_1(\Sigma,B)$ follows the same lines as in the
previous
section. We just give two examples.
\begin{itemize}
\item {\bf 3--torus with its 26 nearest neighbours}: We use the 
numerotation of Fig.~\ref{fig3} and the notation\footnote{This 
labelling of the cells is inspired from the labelling of the principal 
axes in (ordinary) crystallography (see e.g. Phan \cite*{phan90}).} 
$123\equiv g_1\circ g_2\circ g_3 (x)$, $0\equiv x$ etc.  We find that 
the pairs
\begin{itemize}
\item[a:]
$(0,1)$, $(0,2)$ and $(0,3)$ appear 18 times,
\item[b:] $(0,12)$, $(0,15)$,
$(0,23)$, $(0,35)$, $(0,16)$ and $(0,13)$
appear 12 times,
\item[c:]
$(0,123)$, $(0,234)$, $(0,345)$ and $(0,135)$ appear 8 times,
\item[d:]
$(1,4)$, $(2,5)$ and $(3,6)$ appear 9 times,
\item[e:] $(12,45)$,
$(15,24)$, $(13,46)$, $(34,16)$,
$(23,56)$ and 

$(35,26)$ appear 3
times,
\item[f:] $(1,45)$, $(1,24)$, $(2,45)$, $(2,15)$ and 8 other pairs
obtained by
permutations appear 6 times,
\item[g:] $(6,123)$, $(6,234)$,
$(6,345)$, $(6,135)$,
$(1,245)$, $(1,456)$, $(2,456)$, $(2,156)$,
$(4,156)$, $(4,126)$, $(5,246)$

and $(5,126)$ appear 4 times,
\item[h:]
$(24,156)$, $(12,456)$, $(15,246)$, $(45,126)$,
$(13,246)$, $(13,456)$,
$(23,456)$, $(23,156)$, $(43,156)$, $(43,126)$,
$(53,246)$ and $(53,126)$
appear twice,
\item[i:] $(123,456)$ appears once,
\end{itemize}
so that 
$\nu_1(T^3,26)=3\times17+6\times11+4\times7+8\times3+12\times5+ 
12\times3+1 2\times1+1\times0=282$.  The eight families of pairs are 
represented on Fig.~\ref{fig3b}.  In that example about 80\% of the 
increments are in ${\cal N}$.  Again, we recall that when $A=1$, we only 
collect type I pairs.  
\item {\bf Weeks manifold with its 18 nearest 
neighbours}~: This manifold \cite{weeks85} is the smallest known 
compact hyperbolic manifold.  Its description can be found by using 
the software {\it SnapPea} where it has the closed census m003(-3,1).  
Its volume, in units of the curvature radius, is $0.94272$ and its 
fundamental polyhedron has 18 faces (see also Appendix A in Lehoucq et al.
\cite*{lehoucq99}).
It can then be shown after (some tedious computations) that
\begin{equation}
\nu_1(\mathrm{Weeks},18) = 90,
\end{equation}
which implies that ${\cal N}= 9A(11A-1)$.
\end{itemize}

\begin{figure}
\resizebox{\hsize}{!}{\includegraphics{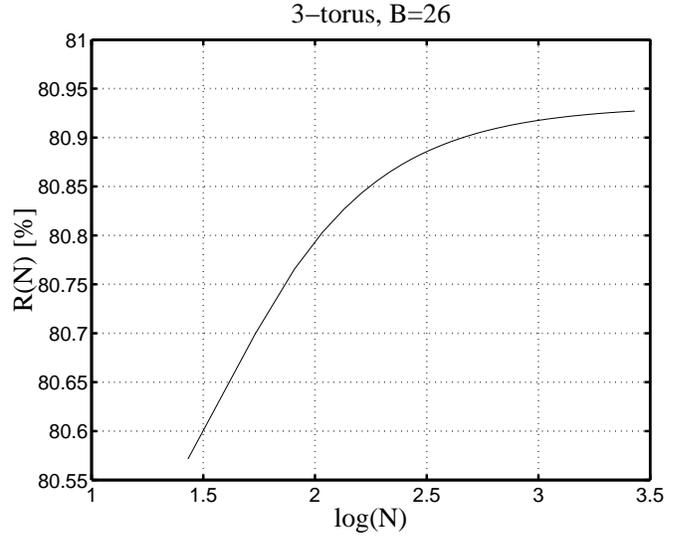}}
\caption{Plot of ${\cal R}(N)$ for the 3--torus $T^3$ with $B=26$ copies.}
\label{figt3}
\end{figure}

\begin{figure}
\resizebox{\hsize}{!}{\includegraphics{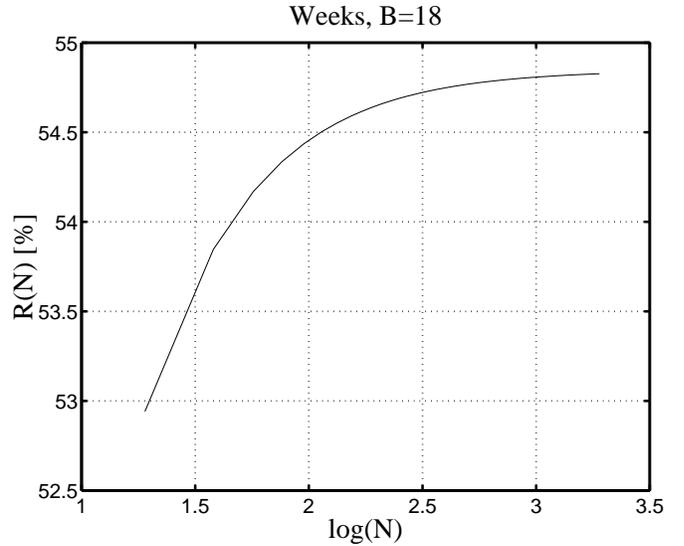}}
\caption{Plot of ${\cal R}(N)$ for the Weeks manifold with $B=18$ copies.}
\label{figweek}
\end{figure}

The CCP--index ${\cal R}$ is given as a function of $\nu_{1}$, $A$ 
(the number of sources in the fundamental domain) and $B$ (the number 
of copies of the domain within the observable universe) by:
\begin{equation}
{\cal R} = \frac{A[(2\nu_1-B)A-B]} {(B+1)^2A^2-(B+1)A-2}.
\end{equation}
In figures \ref{figt3} and \ref{figweek} we plot ${\cal R}$ as a
function of the number of objects $N$ for the two 3--dimensional
examples given above, respectively $\lbrace T^3,26\rbrace$ ($N=27A$) and 
$\lbrace \hbox{Weeks},18\rbrace$ ($N=19A$).

By comparison, when computing numerically the CCP--index in a
simply-connected Friedmann-Lema\^{\i}tre model with the same number of
objects and copies, one gets ${\cal R}(N) \simeq 0$ with some
negligible fluctuations.

Thus, in this idealised method (which neglects the thickness of the
bin $\epsilon$), ${\cal R}$ provides a statistically relevant
indicator for the existence of a non--trivial topology on scales
smaller than the catalog's limit.

\section{Numerical simulations}\label{IV}

In the previous section we have shown how ${\cal R}$ is a good
indicator of the existence of a topological lensing.  We now test our
new method numerically and show how to implement it.  For that
purpose, assuming a locally hyperbolic space, we generate a toy
catalog of cosmic sources as explained in Lehoucq et
al. \cite*{lehoucq99} and we compute the CCP--index.

First, we need to determine the radial distance as a function of 
redshift (since the catalogs provide the angular coordinate and the 
redshift of the sources), which implies knowledge of the 
cosmological parameters.  The local geometry is given by the 
Friedmann-Lema\^\i tre hyperbolic metric
\begin{equation}
\d s^2=-\d t^2+a^2(t)\left(\d \chi^2+\sinh^2{\chi}\d \Omega^2 \right)
\label{METR}
\end{equation}
where $a$ is the scale factor, $t$ the cosmic time and 
$\d \Omega^2\equiv \d \theta^2+\sin^2{\theta}\d \varphi^2$ the infinitesimal 
solid angle.

We compute the relation between the radial coordinate distance 
($\chi$) and the object's redshift $z$.  We start from the Einstein 
equations for the metric (\ref{METR})

\begin{equation}
H^2=\kappa\frac{\rho_m}{3}-\frac{K}{a^2}+\frac{\Lambda}{3},\label{fried1}
\end{equation}
$\rho_m$ being the matter density, $\Lambda$ the cosmological constant 
and $\kappa\equiv 8\pi G/c^4$.  $H$ is the Hubble constant defined by 
$H\equiv \dot a/a$ with $\dot X\equiv \partial_tX$.  We choose the 
units such that the curvature index is $K=-1$.  Introducing 
$\Omega_\Lambda\equiv \Lambda/3H^2$, $\Omega_m\equiv\kappa\rho_m/3H^2$ 
and the redshift $z$ defined by $1+z\equiv a_0/a$, (\ref{fried1}) can 
be rewritten as (see e.g. Peebles \cite*{peebles93})

\begin{equation}
\frac{H^2}{H_0^2}=\Omega_{m0}(1+z)^3+\Omega_{ 
\Lambda0}+(1-\Omega_{m0}- \Omega_{\Lambda0})(1+z)^2.\label{fried2}
\end{equation}
We have used equation (\ref{fried1}) evaluated today (i.e. at $t=t_0$) 
and we have assumed that we were in a matter dominated universe, so 
that $\rho_m\propto a^{-3}$ (this hypothesis is very good since we 
restrict ourselves to small redshifts).

\begin{figure}
\resizebox{\hsize}{!}{\includegraphics{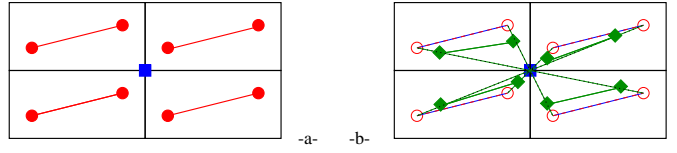}}
\caption{When the cosmological parameters are well estimated,
the redshift--distance relationship enables to reconstruct the
isometries (a).  However, when the estimation is wrong, all the
determination of the radial distances are biased and there is no more
correlated pairs in the coordinate space (b).  The observer is
represented by the square sitting at the centre.}
\label{distw}
\end{figure}

\begin{figure}
\resizebox{\hsize}{!}{\includegraphics{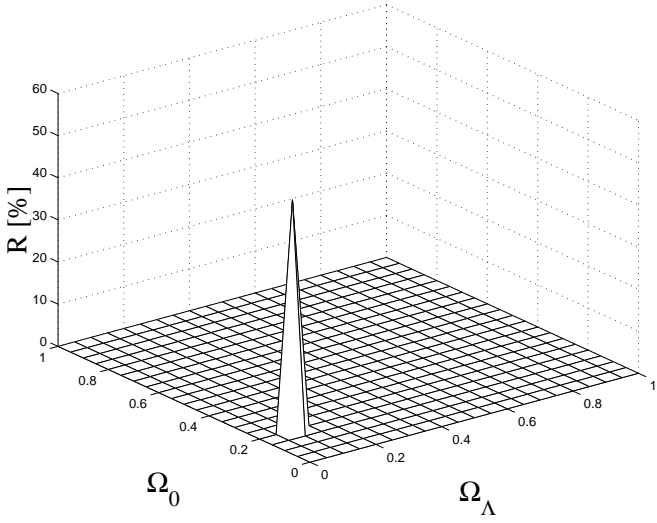}}
\caption{Weeks $A=30$, $\Omega_0=0.2$,
$\Omega_\Lambda=0.1$, $z=3$ (quasars).
We check that the topological signal
stands out only at the 
right values of the cosmological parameters
selected for the simulation.}
\label{sim1}
\end{figure}

The radius of the observable region at a redshift $z$ is given by 
integration of the radial null geodesic equation $d\chi=\d t/a$ and 
reads
\begin{eqnarray}
\chi(z)&\equiv&\int_{a_0}^a\frac{\d a}{a\dot a}\nonumber\\
&=&
\int_{\frac{1}{1+z}}^1\frac{
\sqrt{1-\Omega_{m0}-\Omega_{\Lambda0}}\d x}
{x\,\sqrt{\Omega_{\Lambda0}x^2+(1-\Omega
_{m0}-\Omega_{\Lambda0})
+\frac{\Omega_{m0}}{x}}}.\label{chi1}
\end{eqnarray}
Equation (\ref{chi1}) is integrated numerically and the result can be
compared, when $\Omega_\Lambda=0$, to the analytic expression (see
e.g. Gradstheyn and Ryzhik \cite*{gradstheyn})
\begin{equation}
\chi(z)=\left[\arg\cosh{\left(1+\frac{2(1-\Omega_{m0})}{\Omega_{m0}}x
\right)}\right]^1_{\frac{1}{1+z}}.\label{chi_de_z}
\end{equation}
The determination of the radial distance requires the knowledge of the
cosmological parameters $\Omega_0$ and $\Lambda$ and can be obtained
analytically only when $\Lambda=0$. Thus, if the universe is
multi-connected on sub--horizon scale, the plot of ${\cal R}$ in terms
of $\Omega_0$ and $\Omega_\Lambda$ should exhibit a spike only when the
cosmological parameters have the right value (as shown in
Fig.~\ref{sim1}).  If the cosmological parameters are not exactly
known, the distance (\ref{chi1}) will be ill-determined and the
topological signature will be destroyed (see Fig.~\ref{distw}).

This has two consequences:
\begin{enumerate}
\item One should span the parameters space $(\Omega_0,\Omega_\Lambda)$ 
in order to detect the topological signal, plotting ${\cal 
R}(\Omega_0,\Omega_\Lambda)$.

\item If there is any topological signal, the position of the spike 
gives the values of the cosmological parameters on the scale of the 
catalog's limit (see Fig.~\ref{sim1}).
\end{enumerate}

As a concrete example we proceed as follows. We first generate a
simulated catalog by choosing the number of objects in the fundamental
domain ($A=30$), the topology (Weeks manifold) and the cosmological
parameters (e.g.  $\Omega_0=0.2$, $\Omega_\Lambda=0.1$), and we then
use a second code to apply the test, drawing ${\cal R}$ in terms of
the two cosmological parameters. The result is shown in plot
\ref{sim1}. We see that the method works pretty well in the sense that
there is a spike signalling the presence of a topology and determining
the cosmological parameters. But we also can check that a slight
deviation in the choice of the cosmological parameters makes the spike
to disappear. This effect will be discussed in the next section.

Now, if applied with the required accuracy for the cosmological
parameters, the absence of signature will give the lower bound on the
injectivity radius of the universe (in physical units)
\begin{equation}
\frac{r_{\rm inj}}{3000\,h^{-1}\,\hbox{Mpc}}\geq\int^1_{\frac{1}
{1+z_{\rm
max}}}\frac{\d \ln{x}}
{\sqrt{\Omega_\Lambda
x^2+(1-\Omega_0-\Omega_\Lambda)+
\frac{\Omega_0}{x}}}.
\end{equation}

\section{Working with real data}
\label{V}

\begin{figure}
\resizebox{\hsize}{!}{\includegraphics{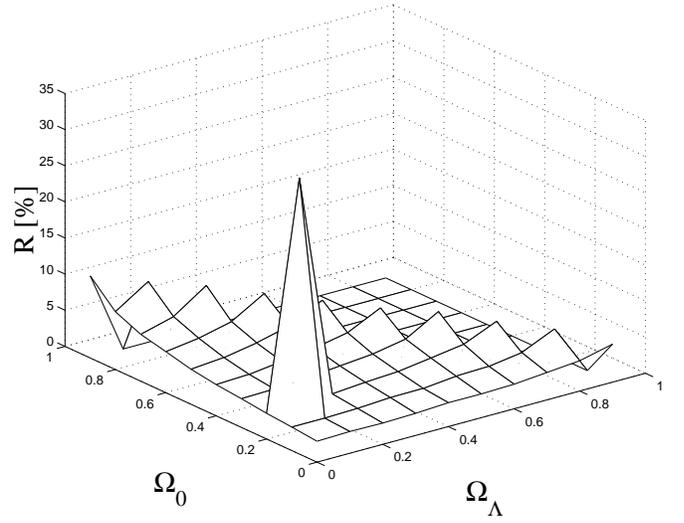}}
\caption{Computation of the CCP--index on a simulated catalog
of depth $z=3$ in a hyperbolic universe model with the Weeks topology,
$\Omega_0=0.3$, $\Omega_\Lambda=0.1$, using a bin width
$\epsilon=10^{-6}$.}
\label{testR}
\end{figure}

When one wants to apply the CCP--method to real data, one faces
a number of problems.  First, we cannot use a zero width bin, one of 
the reasons being that the sources are not exactly comoving.  Let us 
first estimate the precision needed for the cosmological parameters 
when working with a bin of width $\epsilon$ as defined in equation 
(\ref{bin}).  For that purpose, we just assume $\Omega_\Lambda=0$ and 
estimate the precision on $\Omega_0$ from (\ref{chi_de_z}),
\begin{equation}
\left|\frac{\delta\Omega_0}{\Omega_0}\right|= \sqrt{1-\Omega_0} 
\frac{\sqrt{\Omega_0z+1}}{\sqrt{\Omega_0z+1}-1}\epsilon\equiv 
F(\Omega_0,z)\epsilon.
\end{equation}
Assuming $\Omega_0\in[0.2,1[$ and $z\in]0,z_{\rm max}]$, it is easy to 
see that at $\Omega_0$ fixed, $F(\Omega_0,z)$ is a decreasing function 
of $z$ and that $F(\Omega_0,z)\rightarrow\infty$ when $z\rightarrow0$.  
Since $F(\Omega_0,z_{\rm max}=3)\sim 1$, we deduce that
\begin{equation}
\left|\frac{\delta\Omega_0}{\Omega_0}\right|\simeq\epsilon.
\label{accur}
\end{equation}
Indeed, using a catalog with a smaller depth $z_{\rm max}$ will allow 
us to use a smaller resolution for the cosmological parameters.  One 
has thus to find a compromise between depth and resolution.  A deeper 
catalog tests larger topological scales, but requires a better 
accuracy of the cosmological parameters and thus a longer computer 
time.  In Fig.~\ref{testR}, we give an example with a bin width 
$\epsilon=10^{-6}$.  The binning produces a ``background noise'' which 
was absent in Fig.~\ref {sim1}.

\begin{figure}
\resizebox{\hsize}{!}{\includegraphics{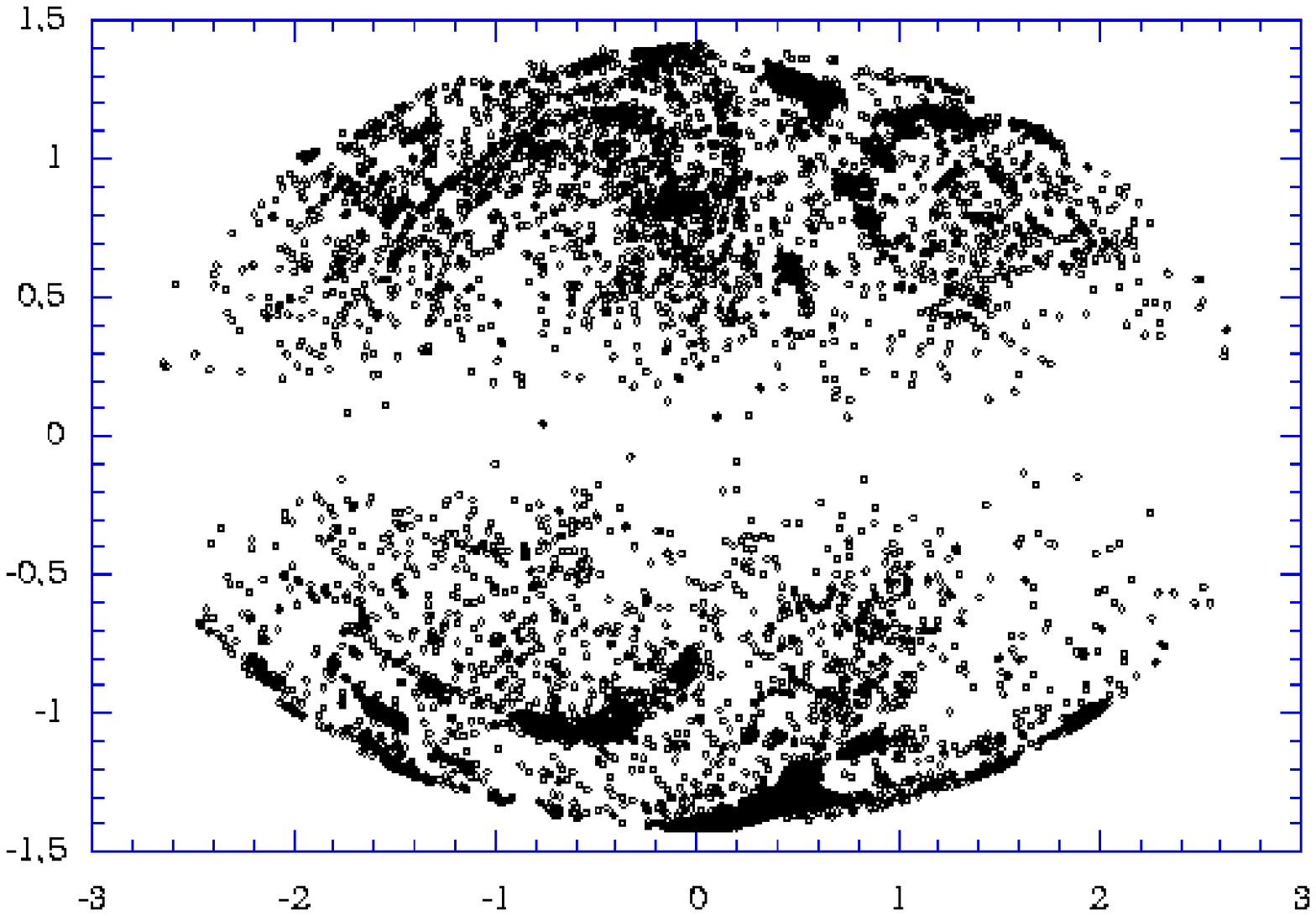}}
\resizebox{\hsize}{!}{\includegraphics{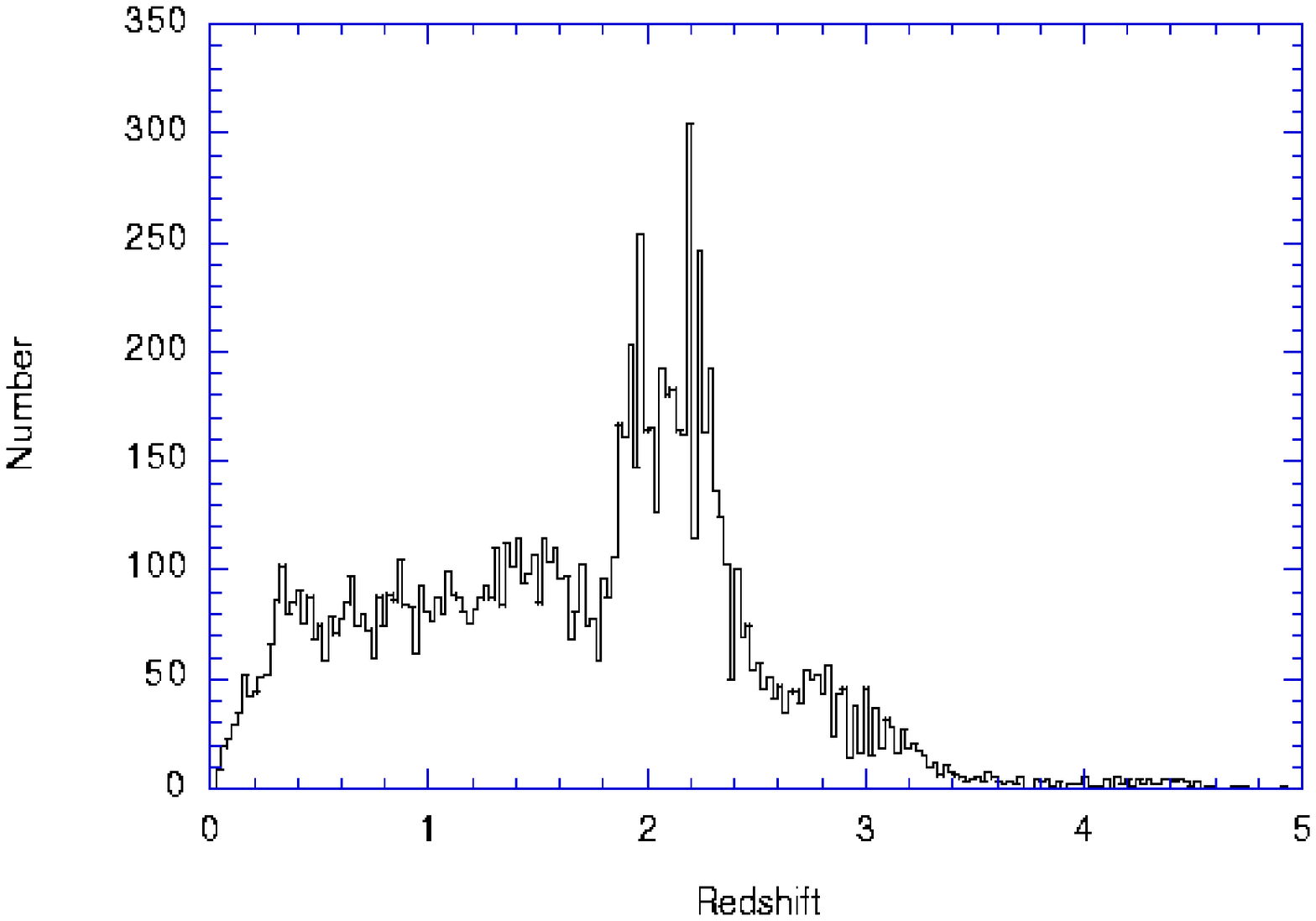}}
\caption{The quasars catalog: (up) the distribution of objects on the
celestial sphere in Hammer-Aitoff equal area projection; (down) the
redshift distribution.}
\label{cat_pic}
\end{figure}

As real data we now consider a quasar catalog\footnote{although quasars are not
as good standard candels as X--ray galaxy clusters, see  Luminet and Roukema \cite*{cargese98}
for a detailed discussion.} \cite{quascat} 
containing 11,301 objects up to a redshift of $z_{\rm max}=4.897$ (but 
only 20\% of the quasars have a redshift greater than 3).  On figures 
\ref{cat_pic} we have depicted its projection on the celestial sphere 
and the redshift distribution of objects.  In Uzan et al.  
\cite*{texas98} the pair separation histogram method, valid only 
in Euclidean spaces, was applied to the same catalog and no topological 
signal was found; this raised the lower bound on the characteristic 
size of Euclidean space to $L\geq 3000\,h^{-1}\hbox{Mpc}$.  This 
limit, corresponding to $L_0/R_H\geq 0.5$, is of the same order as 
the bound $L_0/R_H\geq 0.8$ obtained from the CMB \cite{stevens93}.

We now apply the CCP--method in the Weeks hyperbolic space model
assuming $\Omega_\Lambda=0$ and $\Omega_0$ spanning $[0.2,0.9]$. No
topological signal is found either.  Does it mean that there is no
topological lens effect on scales smaller than $z_{\rm max} \simeq
3$~?  Not necessarily, because we could only apply the test with
precisions $\epsilon=10^{-7}, 10^{-6}, 10^{-5}$, and were unable to
span the cosmological parameter space with the required accuracy
given by (\ref{accur}).  The computer time is one of the main
limitations of our technique.  It depends on the number of sources in
the catalog.  Typically, for the quasar catalog, we had to order about
64 millions pair increments (\ref{incr}) and the run took about 3
minutes (on a SUN Ultra Enterprise 3000 with a 1 Gbytes RAM) for each
couple of cosmological parameters ($\Omega_0$, $\Omega_\Lambda$).
Spanning the full set of reliable values $([0,1] \times [0,1]$ with
their sum smaller than 1) with a given resolution
$\delta\Omega/\Omega$ requires a total running time proportional to
$(\delta\Omega/\Omega)^{-2}$.  Indeed, future observations will
hopefully tighten the range of the cosmological parameters real
values; as a consequence, we will have to span a smaller set of the
cosmological parameters and our simulations will escape the present
computer time limitations.

Now, a second limitation comes from the peculiar velocities of the
sources. Assuming a typical peculiar velocity
$v_p\simeq500\,\hbox{km\,s}^{-1}$ (see e.g. Dekel \cite*{dekel94}) the comoving
position of a given object with respect to the observer is shifted by
the quantity $\alpha(z)$ such that:

\begin{equation}
\alpha(z) \simeq
v_p\frac{\tau(z)}{a_0},
\end{equation}
where $\tau(z)$ is the look-back time. It can be computed following
the same lines as for the computation of (\ref{chi1}), which leads to
\begin{equation}
\frac{\tau(z)}{a_0}=\int_{\frac{1}{1+z}}^1
\frac{\sqrt{1-\Omega_{m0}-\Omega_{\Lambda0}}\d x}
{\sqrt{\Omega_{\Lambda0}x^2+(1-\Omega_{m0}-\Omega_{\Lambda0})
+\frac{\Omega_{m0}}{x}}}.\label{chi2}
\end{equation}

It follows that, since $x\leq1$ in the integrals (\ref{chi1}) and (\ref{chi2}),
\begin{equation}
\frac{\tau(z)}{a_0}<\frac{\chi(z)}{c},
\end{equation}
so that
\begin{equation}
\alpha(z)<\frac{v_p}{c} \chi (z) \simeq 10^{-3}.\chi (z),
\end{equation}
where $\chi (z)$ is the radial distance of the object given by 
(\ref{chi_de_z}).  $\alpha$ is the uncertainty in the comoving 
position of a source {\sl with respect to the observer}.  We show 
below that the corresponding uncertainty induced in the pair 
separation distances is much smaller.

\begin{figure}
\resizebox{\hsize}{!}{\includegraphics{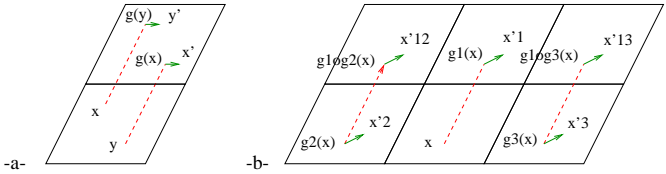}}
\caption{Due to peculiar velocities, two topological images are 
shifted from their strictly comoving position.  We represent here the 
relation between the mathematical images [$g_i(x)$] and the 
topological images [$x'_i$] (a) for $xy$-pairs and (b) for 
$xg(x)$-pairs.}
\label{velocity}
\end{figure}

\begin{itemize}
\item For $xy$--pairs
(see Fig.~\ref{velocity}-a), the space position $x'$ is related to
its strictly comoving position $g(x)$ by
\begin{equation}
x'\simeq g(x)+\frac{\vec v_p(x)}{c}\hbox{dist}[x,g(x)],\label{toto}
\end{equation}

from which we deduce that

\begin{eqnarray}
\hbox{dist}[x',y']&\simeq&
\hbox{dist}[g(x),g(y)]+
\left(\vec{xy}.\vec\nabla\right)
\frac{\vec
v_p(x)}{c}.\frac{\vec{xy}}{\hbox{dist}[x,y]}\nonumber\\
&&\!\!\!\!\!\!\!\!\!\!\!
\!\!\!\!\!+\frac{\vec
v_p(x)}{c}
\frac{\left(\hbox{dist}[x,g(x)]-\hbox{dist}[y,g(y)]
\right)}{\hbox{dist}[x,y]}.
\vec{xy}.
\end{eqnarray}

Since $\hbox{dist}[g(x),g(y] = \hbox{dist}[x,y]$ the uncertainties 
come from two factors
\begin{enumerate}
\item the velocity gradient; however, assuming
large scale velocity flows, two neighbouring images have similar
peculiar velocities and the gradient is very small.

\item a term proportional to the difference of separation distances, 
which vanishes if $g$ is a Clifford translation.  Otherwise (as seen 
in Fig.~11 of Lehoucq et al. \cite*{lehoucq99}), for neighbouring points 
$\hbox{dist}[x,g(x)]-\hbox{dist}[y,g(y)]\ll r_+$, where $r_+$ is the 
largest characteristic size of the manifold [i.e. the radius of the 
smallest geodesic ball containing the fundamental domain, see e.g. 
Fig.~10 in Luminet and Roukema \cite*{cargese98}].
\end{enumerate}
Indeed, we will lose some of the signal from pairs between far apart 
objects.

\item For $xg(x)$--pairs (see
Fig.~\ref{velocity}-b) the position of $g_{i}(x)$ is shifted by
$\vec v_{p}(x)\hbox{dist}[x,g_i(x)]/c$ so that
\begin{eqnarray}
\hbox{dist}[x'_3,x'_{13}]&\simeq&\hbox{dist}[g_3(x),g_3\circ g_1(x)]+
\frac{\vec v_p(x)}{c}.\vec{xg_1(x)}\nonumber\\
&&\!\!\!\!\!\!\!\!
\frac{\left(\hbox{dist}[
x,g_1\circ g_3(x)]-\hbox{dist}[x,g_1(x)]
\right)}
{\hbox{dist}[g_3(x),g_1\circ g_3(x)]}.
\end{eqnarray}

Since $\hbox{dist}[g_3(x),g_3\circ g_1(x)]=\hbox{dist}[x,g_1(x)]$, it 
is easy to see that
\begin{eqnarray}
\hbox{dist}[x'_3,x'_{13}]-\hbox{dist}[x'_2,x'_{12}]\simeq \frac{\vec 
v_p(x)}{c}.\vec{xg_1(x)}\times \nonumber\\
\frac{\left(\hbox{dist}[x,g_2\circ g_ 3(x)]- \hbox{dist}[x,g_1\circ 
g_3(x)] \right)} {\hbox{dist}[x,g_1(x)]}.
\end{eqnarray}
This strictly vanishes if, for instance, $g_2=g_1^{-1}$.  Otherwise, 
$\hbox{dist}[x,g_2\circ g_3(x)]- \hbox{dist}[x,g_1\circ g_3(x)]$ is 
much smaller than $r_+$.  This can be understood from the fact that the 
look-back times of $g_2\circ g_3(x)$ and $g_1\circ g_3(x)$ are of the 
same order and anyway, their difference is much smaller than the 
expected time for a photon to wrap around the universe.
\end{itemize}

To our knowledge, such a discussion about the uncertainty introduced
by proper velocities on the positions of topological images has never
been correctly addressed.  Although the uncertainty induced by proper
velocities is low, it will cause the CCP--index to be smaller than its
theoretical value, since some correlated pairs constructed from far
apart images will be lost.  However, the uncertainty could further be
reduced if, for instance, we were able to obtain a 6--D catalog
including the peculiar velocities, in order to correct the position
from the shift given by (\ref{toto}), assuming that this velocity
field has a weak time dependence. One also has to take into account
the uncertainty in distance determination and spectroscopic errors
(see \cite{roukema96} for a discussion of these effects).

\section{Conclusions}

In this article we have introduced a new method for detecting the
topology of the spatial sections of the universe using 3D--catalogs of
discrete sources. The main motivation for looking for such a method
was the failure of the standard cosmic crystallography for detecting
the topology of locally hyperbolic spaces.

Our method is based on the construction of a {\it CCP--index} as
detailed in Sect.~\ref{III}. The main difference with the cosmic
crystallography is the collecting process of all correlated pairs in
the catalog, which enhances the signal associated to the existence of
a non trivial topology. We then showed the statistical relevance of
this new method both on analytic computations and on numerical
simulations.

Indeed, all the simulations were based on an idealised method applied
to idealised catalogs. We then took into account more realistic
 situations, discussing the effect of the finite width of the bins, of
the peculiar velocities of the objects and of their distance
uncertainties (Sect.~\ref{V}).

To finish, we applied our method to a quasar catalog and found no
topological signature. As explained in Sect.~\ref{V}, this does not
prove that the topological scale of the universe is larger than the
catalog's size because we were limited by computer time in spanning
the full parameter space with the required accuracy.

To achieve this task, we would have to test around $10^{6}$ couples of
cosmological parameters, giving a total computation time on our SUN
workstation greater than five years. A reasonable computation time can
be obtained by increasing the calculation speed by two orders of
magnitude at least. Thus, we are now trying to implement our method on
a Cray T3E parallel computer with 288 DEC Alpha processors in order to
gain a factor of at least 100 in computation time.  It is worth noting that
total scanning of the parameter space can be done by few hundreds
computers since couples of cosmological parameters can be tested
independently.

Indeed, we have to wait for tighter constraints on the cosmological 
parameters coming from various sides of observational cosmology (see 
e.g. the proceedings of the \- XXXIII$^{\rm rd}$ Rencontres de 
Moriond \cite*{moriond98} for a review) in order to reduce the 
parameter space.

Now, it is clear that our method cannot help us determine the exact
topology of the spatial sections of the universe.  It will only
provide a signature of the compactness of these spatial sections on
scales smaller than the catalog depth. Indeed, once this information
is known, one can think of developing a shape recognition code (for
triplets, quadruplets, etc.) that will enable us to reconstruct the
elements of the holonomy group [note that the group structure will
allow us to reject false identifications and to infer the existence of
missing objects in the catalog].

This approach can be thought of as the 3--D analog of the 2--D circles
method using the cosmic microwave background data
\cite{cornish98:a,cornish98:b,cornish98:c}.  The same as the COBE data
does not have the required resolution to exhibit pairs of matched
circles if the universe is small, and the technique also has to wait
for future satellite missions such as MAP and Planck, the CCP--method
is presently limited to 3--D catalogs with low redshifts.  Until now
we are limited to quasars ($z_{\rm max} \simeq 3$).  A recent survey
\cite{chen98} in the Hubble Deep Field south NICMOS field found 17
galaxies with redshifts between 5 and 10 and 5 galaxies with redshifts
above 10, among a total of 323 galaxies.  Such deep surveys can enable
us hope that we may apply our method to deeper catalogs in the future.
Also, just as the circle method will suffer from the degradation in
the circle match caused by detector noise (see e.g. Fig.~7 of Cornish
et al. \cite*{cornish98:c}), our CCP--method suffers from degradation
caused by velocity and distance uncertainties.

To finish, let us extrapolate a little. If the spatial sections of our
universe have any (observationally relevant) topological property, the
graph of ${\cal R}$ in terms of the two parameters ($\Omega_0$,
$\Omega_\Lambda$) should exhibit a resonance at the right value of
these parameters (otherwise, the distance determination being wrong,
no signature could be detected). Besides the detection of a non trivial
topology, the CCP--method can help us determine the cosmological
parameters on the scale of the catalog limit and will, in that case,
be an interesting tool to use together with the other methods
(theoretical and observational) designed to determine these
parameters.  Note that the CCP--method for determining the
cosmological parameters is purely geometric, contrary to the ones
using, e.g., the cosmic microwave background anisotropies which assume a
given scenario of structure formation. Thus, in
the case of multi-connected universe, it will help to improve the
estimation of these parameters from the CMB.  Indeed this is, at the
time being, only prospective.

In conclusion, our new crystallographic method is well backed up
mathematically but suffers from practical implementation difficulties.
It is, however, important to realise that a 3D--catalog of cosmic
sources contains more information than previously suspected, and that
the joint study of topology and cosmology can lead to a better
determination of the curvature parameters.

\begin{acknowledgements}
We thank F. Coelho and B. Pichon for discussion concerning the
numerical implementation of the method and J. Weeks and N.J. Cornish
for their precious comments and suggestions.  J.-P. Uzan also wants to
thank R.  Durrer, P. Peter and F. Vernizzi for discussions during the
redaction of this work and D.A. Steer for her comments on the
manuscript.
\end{acknowledgements}
\bibliographystyle{astron}
\bibliography{topology}
\end{document}